\documentclass[prl,twocolumn,amssymb]{revtex4}
\usepackage{graphicx}
\usepackage{dcolumn}
\usepackage{bm}
\usepackage{wasysym}

\newcommand{\beq}{\begin{equation}}
\newcommand{\eeq}{\end{equation}}
\newcommand{\beqn}{\begin{eqnarray}}
\newcommand{\eeqn}{\end{eqnarray}}

\begin{document}
\title{Experimental observables near a nematic quantum critical point\\ in the pnictide and cuprate
superconductors}
\author{Cenke Xu}
\affiliation{Department of Physics, Harvard University, Cambridge,
MA 02138}
\author{Yang Qi}
\affiliation{Department of Physics, Harvard University, Cambridge,
MA 02138}
\author{Subir Sachdev}
\affiliation{Department of Physics, Harvard University, Cambridge,
MA 02138}

\date{\today}

\begin{abstract}
The newly discovered high temperature superconductor
$\mathrm{SmFeAs(O_{1-x}F_x)}$ shows a clear nematic transition
where the square lattice of Fe ions has a rectangular distortion.
Similar nematic ordering has also been observed in the cuprate superconductors.
We provide a detailed theory of experimental observables near such
a nematic transition: we
calculate the scaling of specific heat, local density of states
(LDOS) and NMR relaxation rate $1/T_1T$.
\end{abstract}
 \maketitle

Rapid and important progress has been made in studies of
Iron-oxypnictides superconductors. Various samples with similar
FeAs plane and rare earths have been synthesized and several
different compounds have shown superconductivity over 50K
\cite{Tb50,Pr50,Nd50,Sm50,Tb3,Sm1}, when the parent compound is
doped by either fluorine or oxygen deficiency. Although many
experimental facts, including the pairing symmetry, are still
under debate, all these different samples share common features: a
tetragonal-monoclinic (orthorhombic) lattice distortion and $(\pi,
0)$ spin density wave (SDW) which commonly exist in the undoped
samples, and ``compete" with superconductivity at finite doping.
The SDW and lattice distortion were first observed in
$\mathrm{LaFeAs(O_{1-x}F_{x})}$ by elastic neutron scattering and
X-ray spectroscopy \cite{La1}. Later on this phenomenon was
confirmed in many other samples with La replaced by Ce \cite{Ce1},
Sm \cite{Sm1,Sm2} and Nd \cite{Ndsdw,Ndsdw2}, and also in oxygen
free materials $\mathrm{BaFe_2As_2}$ \cite{Ba1,Ba2},
$\mathrm{SrFe_2As_2}$ \cite{Sr1,Sr2}, and $\mathrm{CaFe_2As_2}$
\cite{Ca1,Ca2}. In all the samples, at the lattice distortion
temperature $T_{c1}$, the resistivity shows a $\lambda$ shaped
anomaly; therefore the $\lambda$ anomaly of resistivity can be
taken as a measure of the lattice distortion in experiments.

Both lattice distortion and SDW are suppressed under doping, but,
in general, the lattice distortion occurs at higher temperature
than the SDW.
In $\mathrm{SmFeAsO_{1-x}F_x}$,
the lattice distortion and superconductivity coexist in a finite range of
doping; the lattice distortion temperature seems to vanish
within the superconducting phase \cite{Sm1,Sm2}, while the
coexistence between SDW and superconductor was never observed. In
Ref. \cite{xms2008,kivelson2008}, the lattice distortion is
attributed to anisotropic antiferromagnetic correlation between
electrons along $x$ and $y$ directions, without developing long
range SDW. Since this order deforms the electron Fermi surface,
equivalently, it can also be interpreted as electronic nematic
order. This nematic order has Ising symmetry, therefore the
transition temperature is controlled by the intralayer spin
interaction, while the long range SDW is controlled by the
interlayer spin interaction which is much weaker. Therefore the
nematic transition (lattice distortion) occurs at a higher
temperature than the SDW in general, and unless very close to the
critical point, the nematic transition at finite temperature
should belong to the 2d Ising universality class. The distance
between the lattice distortion temperature and the SDW temperature
depends on the anisotropy between $ab$ plane and $c$ axis, which
can be checked by comparing the anisotropy of different samples.

The intimate relation between the structure distortion and SDW
phase proposed by Ref. \cite{xms2008,kivelson2008} has gained
support from recent experiments. It is suggested by detailed
X-ray, neutron and M\"{o}ssbauer spectroscopy studies that both
the lattice distortion transition and the SDW transition of
$\mathrm{LaFeAs(O_{1-x}F_x)}$ are second order
\cite{LaFeAs2ndorder}, where the two transitions occur separately.
However, in $\mathrm{AFe_2As_2}$ with $\mathrm{A = Sr, \ Eu, \ Ba,
\ Ca}$, the structure distortion and SDW occur at the same
temperature, and the structure distortion becomes a first order
transition \cite{Ba3,Sr1,Sr2,Sr3,Ca2} (or a very steep second
order transition \cite{ising}). These results suggest that the SDW
and structure distortion are indeed strongly interacting with each
other, and the structure distortion is probably induced by
magnetism.

We also note that the importance of nematic ordering has also
been discussed recently in the context of the cuprate superconductors \cite{kimnematic,huh2008}.
Our results below are presented in the context of the pnictides, but all
of the scaling properties of the experimental observables apply equally to the cuprates.

\begin{figure}
\includegraphics[width=2.8in]{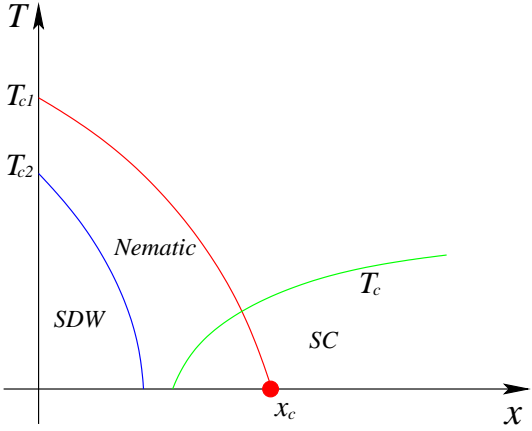}
\caption{The global phase diagram. The blue, red and green curves
are phase boundaries of SDW, nematic (structure distortion) and
superconductivity respectively.} \label{phasedia}
\end{figure}

We focus on the zero temperature nematic phase
transition at finite doping, motivated by the experimental
suggestion of the existence of structure distortion critical point
within the superconducting phase of sample
$\mathrm{SmFeAsO_{1-x}F_x}$ \cite{Sm1,Sm2}. By contrast, the SDW
phase shows no overlap with the superconducting phase in all the
samples studied so far, therefore we will generally ignore it
except for noting that the transition from the SDW to the nematic
order is likely an $z = 1$ O(3) transition, based on the fact that
the SDW order wave vector is independent of doping \cite{Ce1}, so
the low energy particle-hole excitations at the SDW wave vector
vanishes rapidly with small doping and hence make no contribution
to the damping of the SDW order parameter \cite{xms2008}. The
universality class of the nematic transition strongly depends on
the pairing symmetry of the superconducting phase. If it is an
$s-$wave superconductor without nodes, the transition of nematic
order will be an ordinary 3D Ising transition; while if the
superconductor is $d-$wave, the gapless nodal particles may change
the universality class of the nematic transition. The recent STM
\cite{STMSm} and Andreev reflection measurement \cite{andreevSm}
suggest that $\mathrm{SmFeAsO_{1-x}F_x}$ has nodes in the cooper
pair, and the spin susceptibility measured by Knight shift will
tell us whether it is a $p-$wave or $d-$wave pairing. In our
current work we assume a $d-$wave pairing. The universal behavior
of the nematic transition in a $d-$wave superconductor was first
studied in Ref. \cite{kimnematic}. Recently the same theory was
studied carefully, and it was shown that in the infrared limit
there is a special fixed point with logarithmically diverging
velocity anisotropy of the nodal particles \cite{huh2008}. In the
current work we will calculate experimentally relevant quantities
close to this nematic quantum critical point. The global phase
diagram is shown in Fig. \ref{phasedia}.

\begin{figure}
\includegraphics[width=3.0in]{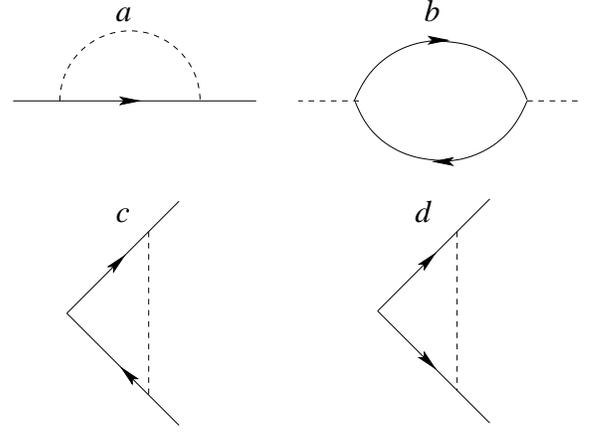}
\caption{The Feynman diagram used in this work, the dashed lines
are the propagators of nematic order parameter $\phi$. $a$, the
self-energy correction to fermion $\Psi$; $b$, the self-energy
correction to $\phi$; $c$, the vertex correction to fermion
bilinear $\Psi^\dagger_i T_A\Psi_j$; $d$, the vertex correction to
fermion bilinear $\Psi^t_i T_A\Psi_j$.} \label{fd}
\end{figure}

The low energy Lagrangian describing the nematic order and nodal
particle reads \cite{kimnematic} \beqn L &=& L_\Psi + L_\phi +
L_{\Psi\phi}, \cr\cr L_\Psi &=& \sum_{a = 1}^{N_f}
\Psi^\dagger_{1a}(\partial_\tau - i v_f
\partial_x \tau^z - i v_\Delta
\partial_y \tau^x)\Psi_{1a}  \cr\cr &+& \Psi^\dagger_{2a}(\partial_\tau - i v_f \partial_y \tau^z -
i v_\Delta
\partial_x \tau^x)\Psi_{2a}, \cr\cr L_\phi &=& \frac{1}{2}(\partial_\tau \phi)^2 +
\frac{c^2}{2}(\nabla\phi)^2 + \frac{r}{2}\phi^2 +
\frac{u_0}{4!}\phi^4, \cr\cr L_{\Psi\phi} &=& \lambda_0\phi
\sum_{a = 1}^{N_f}(\Psi^\dagger_{1a}\tau^x\Psi_{1a} +
\Psi^\dagger_{2a}\tau^x\Psi_{2a}). \label{action}\eeqn The Nambu
fermion $\Psi$ is defined in the standard convention: $\Psi_{1a} =
(f_{1a}, \ \epsilon_{ab}f^\dagger_{3b})$ and $\Psi_{2a} = (f_{2a},
\ \epsilon_{ab}f^\dagger_{4b})$. $a$ and $b$ are spin indices,
$f_1$, $f_2$, $f_3$ and $f_4$ are slow fermion modes at nodal
points $(Q, Q)$, $(- Q, Q)$, $(- Q, - Q)$ and $(Q, -Q)$
respectively. If the system develops long range order of $\phi$,
the four nodal points of the $d-$wave superconductor will be
shifted and break the $C_{4v}$ symmetry down to $C_{2v}$ due to
the coupling $L_{\Psi\phi}$ \cite{kimnematic}. The lagrangian
(\ref{action}) is not Lorentz invariant because in the real system
$v_\Delta/v_f$ is in general not unity. Also, the coupling
$L_{\Psi\phi}$ breaks the Lorentz invariance, since
$\Psi^\dagger\tau^x\Psi$ is only one component of the space-time
current of the Dirac fermion. Therefore a realistic scaling
procedure is to allow $v_\Delta$ and $v_f$ flow independently
under renormalization group (RG).

\begin{figure}
\includegraphics[width=3.3in]{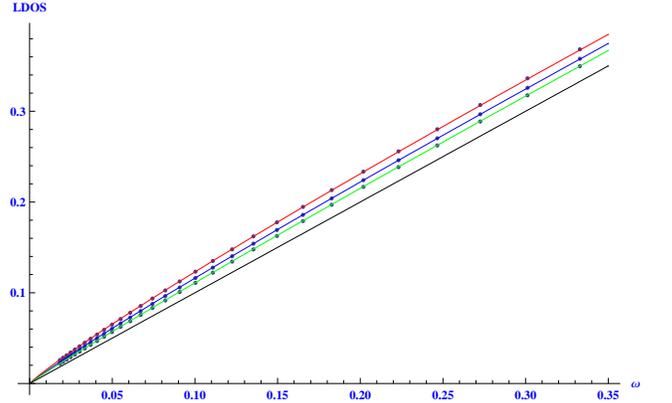}
\caption{Plot of local density of states with $v_{\Delta 0}/v_{f
0} = 1/5$, $v_{\Delta 0}/v_{f 0} = 1/10$, $v_{\Delta 0}/v_{f 0} =
1/20$ from the top to bottom. The horizontal axis is in scale of
$\omega / T_c$. The curves are $y = x^{0.91}$ (red), $y =
x^{0.935}$ (blue), $y = x^{0.955}$ (green), $y = x$ (black). }
\label{ldosplot}
\end{figure}

The nematic transition fixed point in Ref. \cite{huh2008} was
obtained by expansion of $v_\Delta/(v_f N_f)$, assuming a small
initial value $v_{\Delta 0}/v_{f 0}$, and in the current situation
$N_f = 2$. The RG flow of velocities is obtained by calculating
the one-loop self-energy in Fig.  \ref{fd}$a$, with dressed $\phi$
propagator at order $1/N_f$ in Fig . \ref{fd}$b$. The
renormalization condition is chosen to be keeping the coupling
constant $\lambda_0$ in $L_{\Psi\phi}$ invariant under RG flow.
After the one-loop correction, the flow of the self-energy and
velocities reads \beqn \frac{d\Sigma_1}{d\ln \Lambda} &=&
C_1(-i\omega ) + C_2 v_f k_x\tau^z + C_3 v_\Delta k_y\tau^x,
\cr\cr \frac{dv_f}{d\ln \Lambda} &=& (C_1 - C_2) v_f, \cr\cr
\frac{dv_\Delta}{d\ln \Lambda} &=& (C_1 - C_3) v_\Delta, \cr\cr
\frac{d(v_\Delta/v_f)}{d\ln \Lambda} &=& (C_2 - C_3)
(v_\Delta/v_f). \label{velocityRG}\eeqn $\Lambda$ is the momentum
cut-off. $C_1$, $C_2$ and $C_3$ are functions of $v_\Delta/v_f$
and $N_f$, their detailed forms are given in the appendix. Using
these RG equations, we are ready to calculate the scaling of the
local density of states (LDOS) accessible by scanning tunneling
microscope (STM): \beqn \rho(\omega) &\sim&
\int\frac{dk_xdk_y}{(2\pi)^2} \mathrm{Tr} \{
\mathrm{Im}G_{1,ret}(v_fk_x, v_\Delta k_y, \omega) \cr\cr &+&
\mathrm{Im}G_{2,ret}(v_fk_y, v_\Delta k_x, \omega) \} \cr\cr &=&
\frac{1}{v_\Delta v_f} \mathrm{Tr} \{
\int\frac{dk^\prime_xdk^\prime_y}{(2\pi)^2}
\mathrm{Im}G_{1,ret}(k^\prime_x, k^\prime_y, \omega) \cr\cr &+&
\int\frac{dk^\prime_xdk^\prime_y}{(2\pi)^2}
\mathrm{Im}G_{2,ret}(k^\prime_y, k^\prime_x, \omega) \}.
\label{defineLDOS}\eeqn $G_{1}$ and $G_{2}$ are retarded single
particle propagator for $\Psi_1$ and $\Psi_2$. The RG equations in
Eq. \ref{velocityRG} are calculated by rescaling momentum cut-off.
Since $v_\Delta/v_f$ is much smaller than 1 and flows to zero
under RG, for frequency $\omega$, the corresponding momentum scale
is $\tilde{p} = \omega/v_f$. Therefore $d\ln \omega / d\ln
\tilde{p} = 1 + d\ln v_f/ d\ln \tilde{p}$. Now the scaling
equation for $\rho(\omega)$ reads: \beqn \frac{d\ln \rho}{d\ln
\omega} &=& \frac{d\ln \rho}{d\ln \tilde{p}\times \frac{d\ln
\omega}{d\ln \tilde{p}}} = - \frac{d\ln \rho}{d\ln \Lambda \times
(1 - \frac{d\ln v_f}{d\ln \Lambda})} \cr\cr\cr & = & \frac{(1 -
C_1) - \frac{d\ln \left( \frac{1}{v_fv_\Delta}\right)}{d\ln
\Lambda}}{1 - \frac{d\ln v_f}{d\ln \Lambda}} \cr\cr\cr &=& \frac{1
- C_3 - C_2 + C_1}{1+C_2 - C_1}. \label{rgeq2} \eeqn The
ultraviolet cut-off of the theory is taken to be the transition
temperature $T_c$ at the critical doping of nematic transition.
Although $v_\Delta$ will be renormalized to be zero in the
infrared limit, the expansion of $C_2$ and $C_3$ with small
$v_\Delta/v_f$ given in Ref. \cite{huh2008} shows that $v_\Delta$
approaches zero slowly with energy scale, therefore for the
experimentally relevant energy scale, one cannot naively take the
fixed point value of $v_\Delta$. Instead, we have to integrate Eq.
\ref{rgeq2} numerically from the ultraviolet cut-off, and the
result at certain frequency $\omega$ depends on the initial value
of $v_{\Delta 0}/v_{F 0}$. The results of $\rho(\omega)$ with
initial value $v_{\Delta 0}/v_{F 0} = 1/5, 1/10, 1/20$ are plotted
in Fig. \ref{ldosplot}, for frequency between $e^{-4} < \omega/T_c
< e^{-1}$. unlike ordinary $d-$wave superconductor with LDOS
$\rho(\omega) \sim \omega$, in all three plots the LDOS scales
with frequency as \beqn \rho(\omega) \sim \omega^\alpha, \ \
\alpha < 1. \eeqn This simple power law relation fits well for the
frequency range $e^{-4} < \omega/T_c < e^{-1}$, and it can be
checked by STM technique on samples in the quantum critical
regime.

\begin{figure}
\includegraphics[width=3.3in]{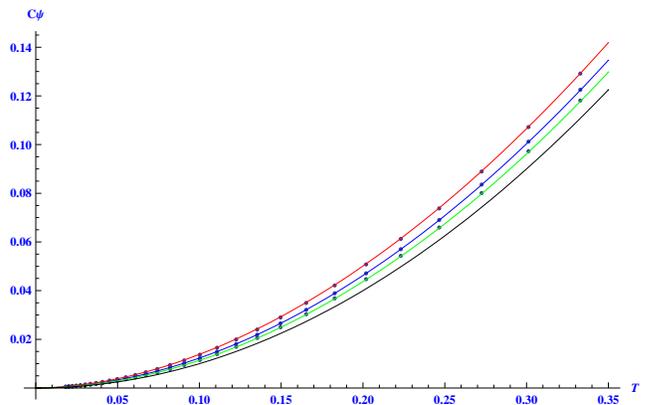}
\caption{Plot of nodal fermion contribution to specific heat with
$v_{\Delta 0}/v_{f 0} = 1/5$, $v_{\Delta 0}/v_{f 0} = 1/10$,
$v_{\Delta 0}/v_{f 0} = 1/20$ from the top to bottom. The
horizontal axis is in scale of $T / T_c$. The curves are $y =
x^{1.86}$ (red), $y = x^{1.91}$ (blue), $y = x^{1.945}$ (green),
$y = x^2$ (black). } \label{cpsiplot}
\end{figure}

The fluctuation of nematic order parameter certainly affects the
thermal dynamical quantities. The free energy $\mathcal{F} = T\ln
\mathcal{Z}/V$ has scaling dimension $d+z$, the singular part of
the free energy can be written as \beqn \mathcal{F} \sim (\xi_\tau
\xi_x \xi_y )^{-1}. \eeqn Now temperature is taken to be the
infrared cut-off: $\xi_\tau \sim 1/T$, and the spatial correlation
length can be estimated to be $\xi_x \sim v_x \xi_\tau$, $\xi_y
\sim v_y \xi_\tau$. The anisotropic velocity of $\Psi_i$ leads to
the following contribution to the free energy: \beqn
\mathcal{F}_\Psi &\sim& \frac{1}{v_fv_\Delta}T^{3}, \cr\cr
\frac{d\ln C_\Psi}{d \ln T} &=& \frac{2 + \frac{d\ln \big(
\frac{1}{v_fv_\Delta}\big)}{d\ln T}}{1 + C_2 - C_1} \cr\cr\cr &=&
\frac{2 + 2C_1 - C_2 - C_3}{1+C_2 - C_1}. \label{rgeqpsi}\eeqn The
$\phi$ contribution to the free energy and specific heat can be
estimated in the same manner, although there is no velocity
anisotropy for the $\phi$ field. The velocity of $\phi$ in the
large $N_f$ limit can be evaluated by calculating the one loop
correction to the self-energy of $\phi$. In the case of small
$v_\Delta/v_f$, the velocity of $\phi$ can be taken to be $v_f$
isotropically: \beqn \mathcal{F}_\phi &\sim& \frac{1}{v_f^2}T^{3}
\cr\cr \frac{d\ln C_\phi}{d \ln T} &=& \frac{ 2+2C_1-2C_2}{1 + C_2
- C_1}. \label{rgeqphi} \eeqn The solutions of Eq. \ref{rgeqpsi}
with different $v_{\Delta 0}/v_{F 0}$ are plotted in Fig.
\ref{cpsiplot}. The equations are solved for the experimentally
relevant temperature range $e^{-4} < T/T_c < e^{-1}$. In general,
the nodal fermions contribute more to the specific heat compared
with $\phi$ field, because $v_\Delta$ scales stronger with
temperature compared with $v_f$. For $e^{-4} < T/T_c < e^{-1}$,
the scaling of specific heat is \beqn C \sim T^\beta, \ \ \beta <
2, \eeqn which distinguishes the current situation from the
ordinary $d-$wave superconductor with nodes.

Another way to probe the density of states is the NMR relaxation
rate $1/T_1$, which is related to the following Green function:
\beqn F(\omega) \sim \int dq_xdq_y \frac{1}{\omega}\chi''(q_x,
q_y, \omega), \eeqn in the limit of $\omega \rightarrow 0$. The
scaling of $1/(T_1T)$ with temperature is the same as the scaling
of $F(\omega)$ with $\omega$, as $T$ and $\omega$ can both serve
as infrared cut-off of the theory. The momentum integrated
susceptibility should involve spin density at various ``slow"
momenta. At the low energy theory of the nodal particles,
following fermion bilinears are low energy spin density modes that
have universal scalings: \beqn \vec{q} &=& (0,0), \ \ \
\Psi^\dagger_1 \sigma^a\Psi_1 + \Psi^\dagger_2 \sigma^a\Psi_2;
\cr\cr \vec{q}_{12A} &=& (2Q, 0), \ \ \
\Psi^\dagger_2\sigma^a\Psi_1; \cr\cr  - \vec{q}_{12A} &=& (-2Q,
0), \ \ \ \Psi^\dagger_1\sigma^a\Psi_2; \cr\cr \vec{q}_{12B} &=&
(0, 2Q), \ \ \ \Psi^t_1\tau^y \sigma^y\sigma^a\Psi_2; \cr\cr -
\vec{q}_{12B} &=& (0, -2Q), \ \ \ \Psi^\dagger_2\tau^y\sigma^a
\sigma^y\Psi^\ast_1, \ \ \ \cr\cr \vec{q}_{11} &=& (2Q, 2Q), \ \ \
\Psi^t_1\tau^y \sigma^y\sigma^a\Psi_1; \cr\cr - \vec{q}_{11} &=&
(-2Q, -2Q), \ \ \ \Psi^\dagger_1\tau^y\sigma^a
\sigma^y\Psi^\ast_1, \cr\cr \vec{q}_{22} &=& (-2Q, 2Q), \ \ \
\Psi^t_2\tau^y \sigma^y\sigma^a\Psi_2; \cr\cr - \vec{q}_{22} &=&
(2Q, -2Q), \ \ \ \Psi^\dagger_2\tau^y\sigma^a \sigma^y\Psi^\ast_2
\label{bilinear}\eeqn $\sigma^a$ are three spin Pauli matrices. To
evaluate $F(\omega)$ we need to calculate the correlation of all
the fermion bilinears above. The susceptibility $\chi$ gains
fermion self-energy correction as in Fig. \ref{fd}$a$ as well as
vertex corrections Fig \ref{fd}$c$ and Fig. \ref{fd}$d$ for
vertices $\Psi^\dagger_i T_A\Psi_j$ and $\Psi^t_i T_A \Psi_j$
respectively. For a general fermion bilinear with flavor matrix
$T_A$, the vertex correction RG equation is conventionally written
as \beqn \frac{d T_A}{d\ln \Lambda} = C_A T_A.
\label{vertexRG}\eeqn $C_A$ is a function of $v_\Delta/v_f$. The
details of calculations are given in the appendix, the results are
\beqn C_{0} &=& - C_1, \ \ \ C_{\tau^x } = C_4 = - C_3, \cr \cr
C_{\tau^y } &=& C_3 - C_1 - C_2, \ \ \ C_{\tau^z} = - C_2 \cr\cr
C_{12A} &=& C_{12B} = - 0.3486 \frac{v_\Delta/v_f}{N_f} + \cdots
\cr \cr C_{11} &=& C_{22} = -C_{\tau^y}. \label{vertices}\eeqn
$C_0$ is the vertex correction for fermion bilinear
$\Psi^\dagger_i\Psi_i$ and $\Psi_i^\dagger \sigma^a\Psi_i$,
because spin is a good quantum number spin Pauli matrices do not
change the vertex correction. $C_{12A}$ and $C_{12B}$ are vertex
corrections to $\Psi^\dagger_1\sigma^a\Psi_2$ and $\Psi^t_1\tau^y
\sigma^y\sigma^a\Psi_2$, $C_{11}$ and $C_{22}$ are vertex
corrections to $\Psi^t_1\tau^y\sigma^y\sigma^a\Psi_1$ and
$\Psi^t_2\tau^y\sigma^y \sigma^a\Psi_2$ respectively.

\begin{figure}
\includegraphics[width=3.3in]{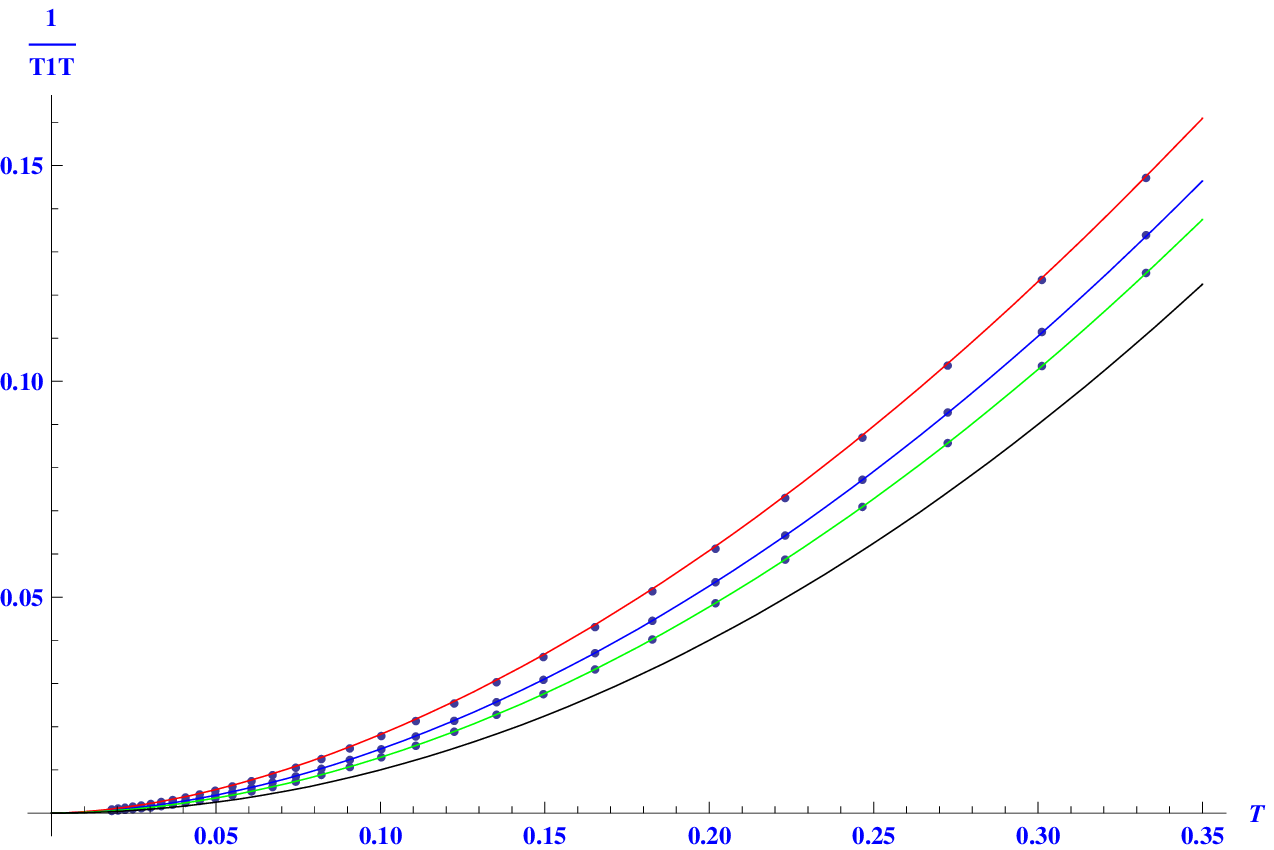}
\caption{Plot of $1/(T_1T)$ with $T$ from contribution of spin
density at $\vec{q} = (0, 0)$, with $v_{\Delta 0}/v_{f 0} = 1/5$,
$v_{\Delta 0}/v_{f 0} = 1/10$, $v_{\Delta 0}/v_{f 0} = 1/20$ from
the top to bottom. The horizontal axis is in scale of $T / T_c$.
The curves are $y = x^{1.74}$ (red), $y = x^{1.83}$ (blue), $y =
x^{1.89}$ (green), $y = x^2$ (black). } \label{00plot}
\end{figure}

\begin{figure}
\includegraphics[width=3.3in]{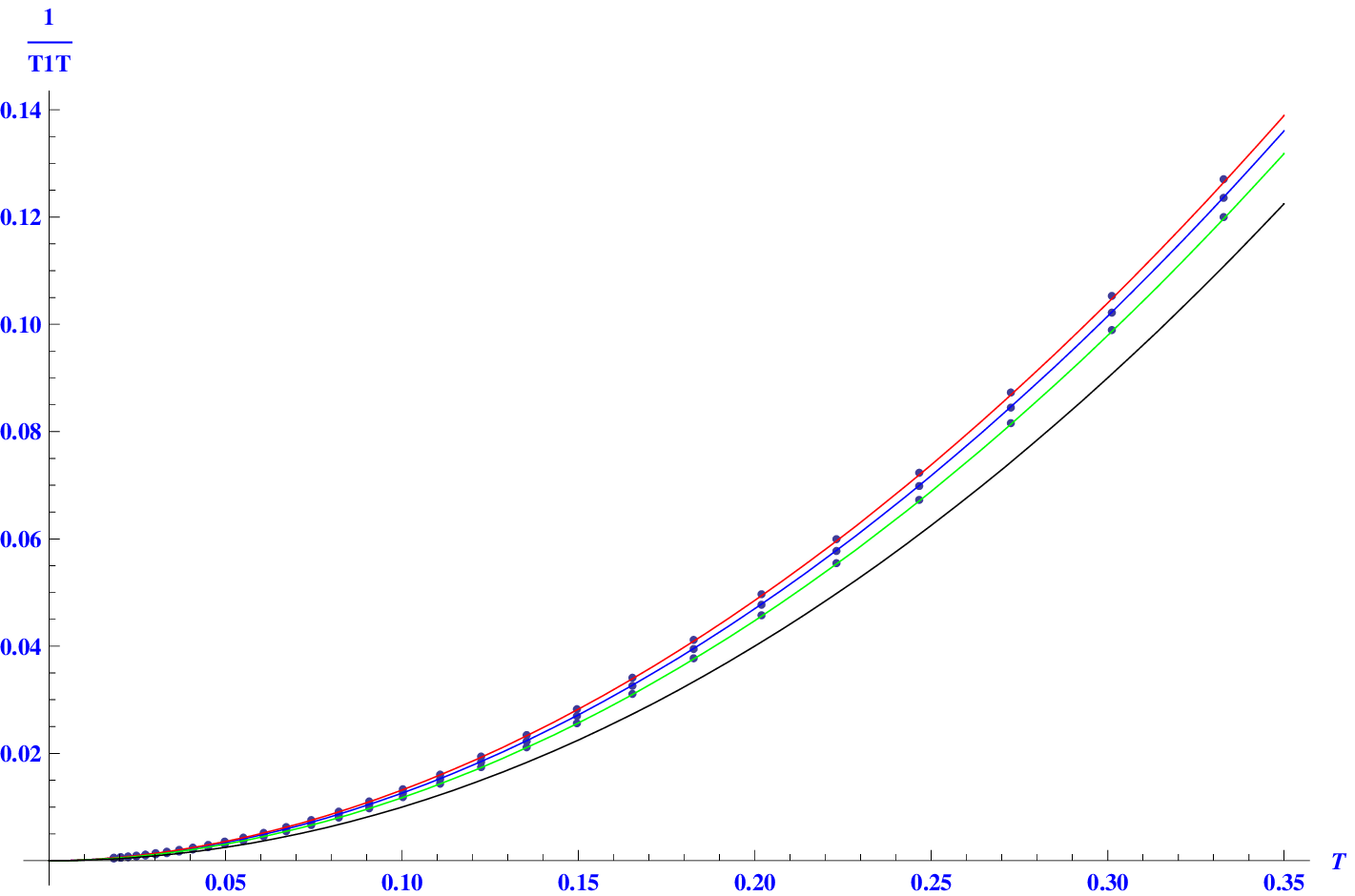}
\caption{Plot of $1/(T_1T)$ with $T$ from contribution of spin
density at $\vec{q} = (2Q, 0)$ and $\vec{q} = (0, 2Q)$, with
$v_{\Delta 0}/v_{f 0} = 1/5$, $v_{\Delta 0}/v_{f 0} = 1/10$,
$v_{\Delta 0}/v_{f 0} = 1/20$ from the top to bottom. The
horizontal axis is in scale of $T / T_c$. The curves are $y =
x^{1.88}$ (red), $y = x^{1.9}$ (blue), $y = x^{1.93}$ (green), $y
= x^2$ (black). } \label{q0plot}
\end{figure}

\begin{figure}
\includegraphics[width=3.3in]{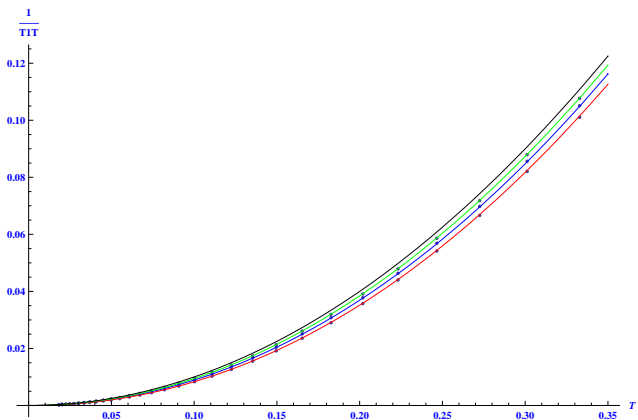}
\caption{Plot of $1/(T_1T)$ with $T$ from contribution of spin
density at $\vec{q} = (2Q, 2Q)$ and $\vec{q} = (2Q, - 2Q)$, with
$v_{\Delta 0}/v_{f 0} = 1/5$, $v_{\Delta 0}/v_{f 0} = 1/10$,
$v_{\Delta 0}/v_{f 0} = 1/20$ from the bottom to the top. The
horizontal axis is in scale of $T / T_c$. The curves are $y =
x^{2.08}$ (red), $y = x^{2.05}$ (blue), $y = x^{2.025}$ (green),
$y = x^2$ (black). } \label{qqplot}
\end{figure}

After taking into account of both self-energy and vertex
corrections, the scaling equation of the relaxation rate reads:
\beqn \frac{d\ln F(\omega)}{d\ln \omega} &=& \frac{2 - 2C_1 - 2C_A
+ 2 \frac{d\ln \big( \frac{1}{v_fv_\Delta}\big)}{d\ln \omega}}{ 1
+ C_2 - C_1} \cr\cr\cr &=& \frac{2 - 2C_A - 2C_2 - 2C_3 + 2C_1}{1
+ C_2 - C_1}. \label{rgeq3} \eeqn Equation (\ref{rgeq3}) leads to
the following scaling equation for $1/(T_1T)$ with temperature:
\beqn \frac{d\ln 1/(T_1T)}{d\ln T}  = \frac{2 - 2C_A - 2C_2 - 2C_3
+ 2C_1}{1 + C_2 - C_1}. \label{rgeq4} \eeqn The contribution from
different fermion bilinear components listed in Eq. \ref{bilinear}
should be solved individually, and the solutions of each component
are plotted in Fig. \ref{00plot}, \ref{q0plot} and \ref{qqplot}.
One can see that at small frequency the spin density at $(0, 0)$
makes the most substantial contribution, which makes the scaling
of $1/(T_1T)$ differ from the ordinary $d-$wave superconductor:
\beqn \frac{1}{T_1T} \sim T^{\gamma}, \ \ \gamma < 2. \eeqn

The superconductivity can be fully suppressed by strong enough
magnetic field. Since the nematic quantum critical point is most
likely in the underdoped regime with low $T_c$, it is possible to
apply strong enough inplane magnetic field $H > H_{c2}$ in
experiments. With superconductivity fully suppressed by inplane
magnetic field, the universality class of the nematic transition
becomes the $z = 3$ theory described in Ref. \cite{xms2008}. The
$z = 3$ quantum critical point leads to a large number of low
energy excitations, which will contribute to thermal dynamics and
transport. The standard mean field theory leads to following
results at low temperature \cite{xms2008}: \beqn C \sim T^{2/3}, \
\ \ \rho \sim T^{4/3}. \eeqn These results will finally crossover
to $d = 3$ scalings close enough to the critical point: \beqn C
\sim T, \ \ \ \rho \sim T^{5/3}. \eeqn Similar $z = 3$ nematic
quantum critical point was discussed in Ref. \cite{kivelson2001}.

In a summary, in this work we computed the scaling of physical
quantities close to a nematic quantum critical point in a $d-$wave
superconducting phase of the newly discovered material
$\mathrm{SmFeAsO_{1-x}F_x}$, motivated by recent experiments on
the polycrystal sample. For the experimentally relevant energy
range, the scaling of LDOS, specific heat, NMR relaxation rate all
deviates from an ordinary $d-$wave superconductor. 
These results also apply, essentially unchanged, to the cuprate
superconductors \cite{kimnematic,huh2008}. This research is
supported by the NSF under grant DMR-0537077.

\appendix{\section{Appendix}}

In this section we shall calculate the vertex corrections to the
fermion bilinears listed in Eq. \ref{bilinear}. For a general
vertex $\Psi_i^\dagger T_A \Psi_j $, the vertex correction is
calculated according to Feynman diagram in Fig. \ref{fd}$c$: \beqn
\mathcal{T_A} = T_A + \frac{1}{N_f} \int
\frac{d^2p}{(2\pi)^2}\frac{d\Omega}{2\pi} \big[\tau^x G_i(\Omega,
p)\cr\cr \times T_A G_j(\Omega, p) \tau^x \big]
\frac{1}{\Gamma_2(\Omega, p)}. \eeqn For a general vertex
$\Psi_i^t T_A \Psi_j $, the vertex correction is calculated
according to Feynman diagram in Fig. \ref{fd}$d$: \beqn
\mathcal{T_A} = T_A + \frac{1}{N_f} \int
\frac{d^2p}{(2\pi)^2}\frac{d\Omega}{2\pi} (-1) \big[\tau^x G_i(
\Omega, p)\cr\cr \times T_A G_j(\Omega, p) \tau^x \big]
\frac{1}{\Gamma_2(\Omega, p)}. \eeqn $\Gamma_2(\Omega, p)$ is the
self-energy of $\phi$ from integrating out fermions: \beqn
\Gamma_2(\omega, p) &=& \Pi_2(k_x, k_y, \omega) + \Pi_2(k_y, k_x,
\omega), \cr\cr \Pi_2(k_x, k_y, \omega) &=&
\frac{1}{16v_fv_\Delta} \frac{\omega^2 + v_f^2k_x^2}{(\omega^2 +
v_f^2k_x^2 + v_\Delta^2k_y^2)^{1/2}}. \eeqn $\Pi_2$ is reminiscent
of the vacuum polarization of the 2+1d QED, with a gauge invariant
form $\Pi_{\mu\nu}(p) \sim p(\delta_{\mu\nu} - p_\mu p_\nu/p^2)$,
and $\Pi_{2} = \Pi_{xx}(p)$.

The RG equation is defined as the change of parameters after
rescaling the cut-off. The cut-off is introduced in a smooth
function $\mathcal{K}(p^2/\Lambda^2)$, with $\mathcal{K}(0) = 1 $
and falls rapidly when $y \rightarrow 1$. Now following the
notation in Ref. \cite{huh2008}, we change the momentum space
integral to cylindrical coordinates $p_\mu = y\Lambda(v_fx,
\cos\theta, \sin\theta)$, and the RG equations in Eq.
\ref{velocityRG} and Eq. \ref{vertexRG} are obtained when the
cut-off is reduced from $\Lambda$ to $\Lambda - d\Lambda$, with
\beqn C_1 &=& - C_0 = \frac{2(v_\Delta/v_f)}{\pi^3 N_f}
\int_{-\infty}^{+\infty} dx \int_0^{2\pi}d\theta \cr\cr &\times&
\frac{(x^2 - \cos^2\theta - (v_\Delta/v_f)^2\sin^2\theta)}{(x^2 +
\cos^2\theta + (v_\Delta/v_f)^2\sin^2\theta)^2}\mathcal{G}(x,
\theta); \cr\cr\cr C_2 &=& - C_{\tau^z} =
\frac{2(v_\Delta/v_f)}{\pi^3 N_f} \int_{-\infty}^{+\infty} dx
\int_0^{2\pi}d\theta \cr\cr &\times& \frac{(- x^2 + \cos^2\theta -
(v_\Delta/v_f)^2\sin^2\theta)}{(x^2 + \cos^2\theta +
(v_\Delta/v_f)^2\sin^2\theta)^2}\mathcal{G}(x, \theta);\cr\cr\cr
C_3 &=& - C_{\tau^x} = \frac{2(v_\Delta/v_f)}{\pi^3 N_f}
\int_{-\infty}^{+\infty} dx \int_0^{2\pi}d\theta \cr\cr &\times&
\frac{( x^2 + \cos^2\theta - (v_\Delta/v_f)^2\sin^2\theta)}{(x^2 +
\cos^2\theta + (v_\Delta/v_f)^2\sin^2\theta)^2}\mathcal{G}(x,
\theta); \cr\cr\cr C_{\tau^y} &=& - C_{11} = -C_{22}  =
\frac{2(v_\Delta/v_f)}{\pi^3 N_f} \int_{-\infty}^{+\infty} dx
\int_0^{2\pi}d\theta \cr\cr &\times& \frac{(x^2 + \cos^2\theta +
(v_\Delta/v_f)^2\sin^2\theta)}{(x^2 + \cos^2\theta +
(v_\Delta/v_f)^2\sin^2\theta)^2}\mathcal{G}(x, \theta) \cr\cr &=&
C_3 - C_1 - C_2 ; \cr\cr\cr C_{12A} &=& C_{12B} =
\frac{2(v_\Delta/v_f)}{\pi^3 N_f} \int_{-\infty}^{+\infty} dx
\int_0^{2\pi}d\theta \cr\cr (-x^2) &\times& \frac{1}{x^2 +
\cos^2\theta + (v_\Delta/v_f)^2\sin^2\theta} \cr\cr &\times&
\frac{1}{x^2 +(v_\Delta/v_f)^2 \cos^2\theta +
\sin^2\theta}\mathcal{G}(x, \theta) ; \cr\cr\cr
\mathcal{G}^{-1}(x, \theta) &=& \frac{x^2 +
\sin^2\theta}{\sqrt{x^2 + (v_\Delta/v_f)^2 \cos^2\theta +
\sin^2\theta}} \cr\cr &+& \frac{x^2 + \cos^2\theta}{\sqrt{x^2 +
(v_\Delta/v_f)^2 \sin^2\theta + \cos^2\theta}}.  \eeqn

It was noted in Ref. \cite{huh2008} that function $C_3$ has a
rather singular form in the small $v_\Delta/v_f$ limit: \beqn C_3
= \frac{8}{\pi^2}\big[\ln (v_f/v_\Delta) -
0.9601\big]\frac{v_\Delta/v_f}{N_f} + \cdots. \eeqn Plugging this
function into Eq. \ref{velocityRG}, one can see that
$v_\Delta/v_f$ approaches zero a little faster than the ordinary
marginally irrelevant operators, but for the experimentally
relevant energy scale, $v_\Delta/v_f$ still flows slowly.
Therefore all the plots in our paper, though integrated from a
complicated equation, can be fit with a simple power law.

\end{document}